\documentclass[aps,pre,showpacs,onecolumn]{revtex4-1}
\usepackage{graphicx}
\usepackage{graphicx}
\usepackage{amsmath}
\usepackage{amsfonts}
\usepackage{xcolor}
\begin{document}
\title{Synchronization by external periodic force in ensembles of globally coupled phase oscillators}

%\title{Higher-order coupling induces Cyclope states in repulsive Kuramoto networks}

%\title{Cyclope states in repulsive Kuramoto networks: the role of high-order coupling}
%\title{Anti-resonance in switching systems}

\author{Semyon S. Abramov$^{1}$, Maxim I. Bolotov$^{1}$\footnote{Corresponding author, e-mail: maksim.bolotov@itmm.unn.ru}, and Lev A. Smirnov$^{1}$}

\address{$^1$Department of Control Theory, Lobachevsky State University of Nizhny Novgorod,
	23 Gagarin Avenue, Nizhny Novgorod, 603022, Russia}	
\begin{abstract}
We consider the effect of an external periodic force on chimera state in the phase oscillator model proposed by Abrams et al. \cite{1}.
%[Phys. Rev. Lett, v. 101, 00319007 (2008)]. 
Using the Ott--Antonsen reduction the dynamical equations for the global order parameter characterizing the degree of synchronization are constructed. The regions of the global order parameter frequency locking by an external periodic force are constructed. The possibility of stable chimeras synchronization and unstable chimeras stabilization are established.
\end{abstract}

\date{\today}
\draft \maketitle

\section{Introduction} 
There are many models of nonlinear dynamics, one such model is ensembles of phase oscillators. It allows describing multicomponent self-oscillating systems with weak coupling consisting of many elements \cite{9, 15, 21}. Phase reduction is one of the main methods for the study of systems consisting of a large number of oscillating elements. It allows us to understand and describe a wide range of fundamental effects that arise in such systems. In the framework of the phase approximation, the states of the system elements are parameterized on the basis of an angular variable -- the phase. This allows us to consider individual elements as oscillators that oscillate around their limit cycle. When the interaction between elements is weak, the phase approximation can be applied to analyze and predict the behavior of the system as a whole. When considering the effects of synchronization in populations and medias of phase oscillators, the Kuramoto model and its various modifications have a special role, when the coupling between the elements is organized due to the phase difference~\cite{16, 3, 17}. In the work~\cite{1} Abrams and Strogatz proposed a model of two ensembles of globally coupled phase oscillators based on the Kuramoto system. In this case the interaction force between the elements of one ensemble differs from the interaction force between the oscillators in different ensembles. In this model, a chimera state characterized by the coexistence of two groups of oscillators in a system of identical elements: fully synchronous and partially synchronous, was found. This mode is of interest for research and was found both in theoretical works~\cite{2, 13, 23, 11} and in experiments~\cite{19, 10, 20, 7}. One of the main methods in the analysis of the synchronization degree in a system of phase oscillators is the Ott--Antonsen reduction~\cite{12, 14, 4}. This approach allows us to pass from the phase description of ensembles to dynamical equations with respect to the global order parameter. Using the Ott--Antonsen method reduced equations were obtained for both lumped~\cite{12} and distributed systems \cite{14}, including in the presence of an external force~\cite{4}.

The model of a globally coupled phase oscillators with an external periodic force was first considered by Sakaguchi in~\cite{18}. It can be used to describe the circadian rhythm. The circadian rhythm in mammals is regulated by the suprachiasmatic nucleus, which is a population of neurons in an oscillatory mode that is influenced through the optic nerve throughout the day. In~\cite{22} the Ott--Antonsen equations were derived for a system consisting of one ensemble of globally coupled Kuramoto--Sakaguchi phase oscillators with an external periodic force. In this paper~\cite{5} the effect of external periodic force on chimera states was investigated in a spatially distributed system of Kuramoto--Battogktokh phase oscillators. Arnold tongues on the plane amplitude-frequency of external force parameters were obtained, as well as the possibility of stabilization and regularization of chimera state was established. 

This paper considers the effect of an external periodic force on a system of two ensembles of globally coupled oscillators described by the model of Abrams et al.~\cite{1}. Such networks of oscillators, consisting of several globally coupled populations, have a key role in understanding of neural synchronization structures~\cite{6}.

The paper is organized as follows. In Section~\ref{sec:2}, a model of phase oscillators ensembles under the influence of an external periodic force is described. Then the transition to dynamical equations with respect to global order parameters by means of the Ott--Antonsen reduction is demonstrated. The main structures in the phase space of the reduced system, which determine the observed dynamical modes in the phase element system, are considered. The description of the stable stationary chimeras synchronization of and unstable chimera regimes stabilization under the action of an external periodic force are presented in the Section~\ref{sec:3}. In the Summary the main results of the research are formulated.

\section{Model of phase oscillators ensembles}
\label{sec:2}
\subsection{Description of the phase model}
In this paper we study the effects of an external periodic force on chimera modes that are realized in a system of two ensembles of globally coupled identical phase oscillators. We consider a system of $N_1$ and $N_2$ elements in the ensembles, respectively, which is affected by an external periodic force with amplitude $\varepsilon$ and frequency $\Omega$:
\begin{equation}
	\label{eq1}
	\begin{gathered}
		\dot{\varphi}_n^{(1)}\!=\!\omega\!+\!\frac{\mu_1}{N_1}\!\sum\limits_{m=1}^{N_1}\sin(\varphi_{m}^{(1)}\!-\!\varphi_{n}^{(1)}\!-\!\alpha)\!+\!\frac{\mu_2}{N_2}\!\sum\limits_{m=1}^{N_2} \sin(\varphi_{m}^{(2)}\!-\!\varphi_{n}^{(1)}\!-\!\alpha)\!+\!\varepsilon\sin(\Omega t\!-\!\varphi_{n}^{(1)}),\\
		\dot{\varphi}_n^{(2)}\!=\!\omega\!+\!\frac{\mu_1}{N_2}\!\sum\limits_{m=1}^{N_2}\sin(\varphi_{m}^{(2)}\!-\!\varphi_{n}^{(2)}\!-\!\alpha)\!+\!\frac{\mu_2}{N_1}\!\sum\limits_{m=1}^{N_1} \sin(\varphi_{m}^{(1)}\!-\!\varphi_{n}^{(2)}\!-\!\alpha)\!+\!\varepsilon \sin(\Omega t\!-\!\varphi_{n}^{(2)}),
	\end{gathered}
\end{equation}
where $\varphi_{n}^{(m)}$ is the phase of the $n$-th oscillator in the $m$-th ensemble (ensemble number $m=1, 2$), $\omega$ is the natural frequency, $\alpha$ is the phase shift (Sakaguchi parameter), ${\mu_1}$ is the strength of inter-element interaction within ensembles, ${\mu_2}$ is the strength of inter-element interaction between ensembles, $N_{m}$ is the number of elements in the $m$-th ensemble. Thus, the model describes the interaction of each element of one ensemble with other elements in its ensemble with one strength of coupling, while with another ensemble elements the coupling strength is different. Note that the model~\eqref{eq1}, when transferred to a coordinate system uniformly rotating with frequency $\Omega$, is a system of coupled active rotators that allows us to describe the dynamics of the neuron-like elements~\cite{8}.

\subsection{Ott--Antonsen equations}
Ott--Antonsen reduction is often used to study dynamical systems with a large number of coupled oscillators; this method allows us to significantly reduce the number of considered equations and to follow the dynamics not of each element of the system, but of the mean field --- the global order parameter formed by the each population elements.

First of all, let us introduce a rotating coordinate system using the following substitution:
\begin{equation}
	\psi_n^{(1)} = \varphi_n^{(1)} - \Omega t ,\quad \psi_n^{(2)} = \varphi_n^{(2)} - \Omega t,  
\end{equation}
which free the equations~\eqref{eq1} from explicit dependence on time and allow to study the system with periodic force as an autonomous system. The equations~\eqref{eq1} take the following form:
\begin{equation}\label{eq2}
	\begin{gathered}
		\dot{\psi}_n^{(1)}\!=\!(\omega\!-\!\Omega)\!+\!\frac{\mu_1}{N_1} \!\sum\limits_{m=1}^{N_1}\sin(\psi_{m}^{(1)}\!-\!\psi_{n}^{(1)}\!-\!\alpha)\!+\!\frac{\mu_2}{N_2}\!\sum\limits_{m=1}^{N_2}\sin(\psi_{m}^{(2)}\!-\!\psi_{n}^{(1)}\!-\!\alpha)\!-\!\varepsilon \sin \psi_{n}^{(1)},\\
		\dot{\psi}_n^{(2)}\!=\!(\omega\!-\!\Omega)\!+\!\frac{\mu_1}{N_2}  \sum\limits_{m=1}^{N_2}\sin(\psi_{m}^{(2)}\!-\!\psi_{n}^{(2)}\!-\!\alpha)\!+\!\frac{\mu_2}{N_1}\sum\limits_{m=1}^{N_1} \sin(\psi_{m}^{(1)}\!-\!\psi_{n}^{(2)}\!-\!\alpha)\!-\!\varepsilon \sin \psi_{n}^{(2)}.
	\end{gathered}
\end{equation}

To describe the collective interaction in the system~\eqref{eq2}, we introduce a complex global order parameter $R_{m}(t)$ for each ensembles:
\begin{equation}\label{eq3}
	R_{m}(t) = \frac{1}{N_{m}}\sum\limits_{n=1}^{N_{m}} e^{i \phi_n^{(m)}}=\rho_{m}(t)e^{-i\theta_{m}(t)},   
\end{equation}
whose amplitudes satisfy the inequality $\rho_{m} \leq 1$. In the case $\rho_{m}=1$ all oscillators in the $m$-th ensemble are phase synchronized. If the condition $0<\rho_{m}<1$ is fulfilled, we can say that the partial synchronization mode is observed, while in the case $\rho_{m} = 0$ the mode is completely asynchronous. We can write down the dynamic equations with respect to the amplitude and phase of the order parameters. This is easily done by writing the system~\eqref{eq2} in complex variables and substituting~\eqref{eq3}:
\begin{equation}\label{eq4}
	\begin{gathered}
		\dot{\psi}_n^{(1)} = (\omega - \Omega) + \text{Im}[(\mu_1 R_1 + \mu_2 R_2 + \varepsilon e^{i\alpha}) e^{-i(\psi_n^{(1)} +\alpha)}],\\
		\dot{\psi}_n^{(1)} = (\omega - \Omega) + \text{Im}[(\mu_1 R_2 + \mu_2 R_1 + \varepsilon e^{i\alpha}) e^{-i(\psi_n^{(2)} +\alpha)}].
	\end{gathered}
\end{equation}
Let us introduce the functions $H_1(t)$ and $H_2(t)$, which describe the complex fields generated by the dynamics of the ensemble elements acting on each oscillator in the first and second ensemble, respectively:
\begin{equation}\label{eq5}
	\begin{gathered}
		H_1(t) = \mu_1 R_1(t) + \mu_2 R_2(t) + \varepsilon e^{i\alpha},\\
		H_2(t) = \mu_1 R_2(t) + \mu_2 R_1(t) + \varepsilon e^{i\alpha}.
	\end{gathered}
\end{equation}
Then the system~\eqref{eq4} can be represented as:
\begin{equation}\label{eq6}
	\begin{gathered}
		\dot{\psi}_n^{(1)} = (\omega - \Omega) + \text{Im}[H_1(t) e^{-i(\psi_n^{(1)} + \alpha)}],\\
		\dot{\psi}_n^{(2)} = (\omega - \Omega) + \text{Im}[H_2(t) e^{-i(\psi_n^{(2)} + \alpha)}].
	\end{gathered}
\end{equation}

In the thermodynamic limit $N_m \to \infty$, the system \eqref{eq6} can be characterized by the distribution densities $f_{m}(\psi^{(m)},t)$ of the phases $\psi$ at fixed $t$. These functions satisfy the continuity equation
\begin{equation}\label{eq7}
	\frac{\partial f_{m}}{\partial t} + \frac{\partial}{\partial \psi^{(m)}}\Big(\big[(\omega - \Omega) + \text{Im}[H_m(t) e^{-i(\psi^{(m)} + \alpha)}]\big] f_{m}\Big) = 0. 
\end{equation}
Using the approach proposed by Ott and Antonsen \cite{12}, based on the existence of a manifold invariant with respect to the dynamics of the system under study, and due to the periodicity in phase of $\psi^{(m)}$ of the probability density $f_{m}(\psi^{(m)},t)$ the solution of the equation \eqref{eq7} can be sought in the form of the following Fourier series:
\begin{equation}\label{eq8}
	f_{(m)}(\psi^{(m)},t) = \frac{1}{2\pi}\Big(1+\sum\limits_{k=1}^{\infty}(a_{m}^k (t)e^{ik\psi^{(n)}} + \text{c.c.})\Big).
\end{equation}
In~\cite{12,4} it is shown that substituting $a_{m}(t) = R_{m}^{*}(t)$ gives an exact solution of the Eq.~\eqref{eq8} if the global order parameters $R_m(t)$ satisfy the following system:
\begin{equation}\label{eq9}
	\begin{gathered}
		\dot{R}_1 = i(\omega - \Omega)R_1 + \frac{1}{2}(e^{-i\alpha}H_1 - e^{i\alpha}H_1^* R_1^2),\\
		\dot{R}_2 = i(\omega - \Omega)R_2 + \frac{1}{2}(e^{-i\alpha}H_2 - e^{i\alpha}H_2^* R_2^2),
	\end{gathered}
\end{equation}
where the index $*$ stands for the complex conjugation. Then from the system~\eqref{eq9} one can already express the amplitudes $\rho_m$ and phases $\theta_m$ of the complex order parameters $R_{m}(t)=\rho_m(t) e^{i \theta_m(t)}$:
\begin{equation}\label{eq10}
	\begin{gathered}
		\dot{\rho}_1 = \frac{1 - \rho_1^2}{2}(\mu_1\rho_1\cos{\alpha} + \mu_2\rho_2\cos{(\theta_2 - \theta_1 - \alpha)} + \varepsilon\cos{\theta_1}),\\
		\dot{\rho}_2 = \frac{1 - \rho_2^2}{2}(\mu_1\rho_2\cos{\alpha} + \mu_2\rho_1\cos{(\theta_1 - \theta_2 - \alpha)} + \varepsilon\cos{\theta_2}),\\
		\dot{\theta}_1 = (\omega - \Omega) - \frac{\rho_1^2+1}{2\rho_1}(\mu_1\rho_1\sin{\alpha} + \mu_2\rho_2\sin{(\theta_1 - \theta_2 + \alpha)} + \varepsilon\sin{\theta_1}),\\
		\dot{\theta}_2 = (\omega - \Omega) - \frac{\rho_2^2+1}{2\rho_2}(\mu_1\rho_2\sin{\alpha} + \mu_2\rho_1\sin{(\theta_2 - \theta_1 + \alpha)} + \varepsilon\sin{\theta_2}).
	\end{gathered}
\end{equation}
Thus, we obtained a system of reduced equations~\eqref{eq10} describing the modules and phases of the global order parameters. The equilibrium states in this system correspond to stationary states with respect to the order parameter of the phase systems, and periodic motions to the modes of phase dynamics when the order parameter modulus is a periodic function.

\subsection{Dynamics in the absence of external force. Chimera states}
Next, we interest in the chimera state in the system~\eqref{eq1}. In this case, the elements of the first ensemble form a fully phase-synchronous cluster with $\rho_1 = 1$, while the elements of the second ensemble are partially synchronous with $\rho_2 < 1$~\cite{1,2}.
In~\cite{1} it is shown that in the case of no external force at $\varepsilon = 0$ the system~\eqref{eq10} can be reduced to a third-order system in the case of invariance with respect to phase shift. 
Whence, restricting ourselves to the chimera regime, we can pass to a system of two equations with respect to $r = \rho_2$ and $\psi = \varphi_1 - \varphi_2$:
\begin{equation}\label{eq12}
	\begin{gathered}
		\dot{r} = \frac{1-r^2}{2}\Big(\mu r \cos{\alpha} + \nu \cos{(\psi - \alpha)}\Big),\\
		\dot{\psi} = \frac{1+r^2}{2r}\Big(\mu r \sin{\alpha} - \nu \sin{(\psi - \alpha)}\Big) - \mu \sin{\alpha} - \nu r \sin{(\psi + \alpha)}.
	\end{gathered}
\end{equation}
Three equilibrium states can exist in the phase space of the system \eqref{eq12}: (i) the absolutely stable (or unstable) $O_1(r_{1},\psi_{1})$ and (ii) the saddle $O_2(r_2,\psi_{2})$, corresponding to chimera state, as well as (iii) the absolutely stable $O_3(r_3,\psi_{3})$, which corresponds to the fully synchronous regime.
\begin{figure}[!ht]
	\centering{\includegraphics[width=1\linewidth]{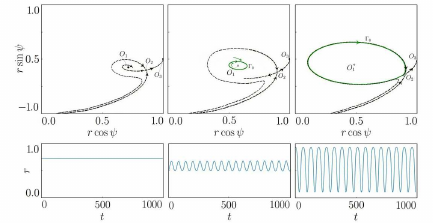}}
	\caption{(a,b,c) Phase portraits of the system~\eqref{eq12} at $\alpha = \pi / 2 - 0.1$. Points $O_1$, $O_2$, $O_3$ are equilibrium states; filled points denote stable modes, hollow points denote unstable modes. The green closed curve $\Gamma_0$ is the limit cycle. The black dashed curves denote the separatrices of the saddle $O_2$. (d,e,f) Dynamics of the variable $r(t)$ corresponding to (a) stationary chimera regime; (b) weakly nonlinear breather chimera state; (c) strongly nonlinear breathing chimera state. Parameters: (a,d) $\mu_1=0.6$, $\mu_2=0.4$; (b,e) $\mu_1=0.64$, $\mu_2=0.36$; (c,f) $\mu_1=0.675$, $\mu_2=0.325$.}
	\label{ris:phase}
\end{figure}

Let us introduce a new parameter $A = \mu_1 - \mu_2$ characterizing the disorder between the forces of intra- and inter-ensemble interaction. When the parameter $A$ is changed from the stable equilibrium state $(r_{1},\psi_{1})$ (Fig. 1a), a stable limit cycle $\Gamma_0$ (Fig. \ref{ris:phase}b) is born as a result of the Andronov--Hopf supercritical bifurcation, which corresponds to the breather chimera state. The breather chimera regime in the system~\eqref{eq1} is a chimera state where the global order parameter amplitude of the partially synchronous cluster exhibits periodic dynamics (Fig. \ref{ris:phase}e). As the disorder of the coupling forces $A$ increases, the limit cycle approaches the saddle separatrices $(r_{2},\psi_{2})$ (Fig. 1c), while the modulus of the order parameter exhibits oscillations with a more nonlinear character (Fig. 1f).

\subsection{Mean-field dynamics of chimeras under external forcing}
Let us further consider the effect of an external periodic force on the chimera modes corresponding to the situations presented in Fig.~\ref{ris:phase}.
In the presence of a periodic external force for a chimera state with $\rho_1 = 1$ and $\rho_2 = r$, the system \eqref{eq10} is reduced to the next third-order system:
\begin{equation}\label{eq13}
	\begin{gathered}
		\dot{r} = \frac{1 - r^2}{2}(\mu_1 r \cos{\alpha} + \mu_2\cos{(\theta_1 - \theta_2 - \alpha)} + \varepsilon\cos{\theta_2}),\\
		\dot{\theta}_1 = (\omega - \Omega) - (\mu_1\sin{\alpha} + \mu_2\rho_2\sin{(\theta_1 - \theta_2 + \alpha)} + \varepsilon\sin{\theta_1}),\\
		\dot{\theta}_2 = (\omega - \Omega) - \frac{\rho_2^2+1}{2\rho_2}(\mu_1\rho_2\sin{\alpha} + \mu_2\sin{(\theta_2 - \theta_1 + \alpha)} + \varepsilon\sin{\theta_2}).
	\end{gathered}
\end{equation}
The equilibrium states $O_k(r^{(k)}, \theta^{(k)}_1, \theta^{(k)}_2)$ ($k=1, 2, \dots$) in the system \eqref{eq13}, where $r^{(k)} \neq 1$, define chimera modes in the original system of phase equations~\eqref{eq1}. When $\varepsilon = 0$ the equilibrium states $(r^{(k)}, \theta^{(k)}_1, \theta^{(k)}_1, \theta^{(k)}_2)$ become degenerate when there exists an entire manifold of equilibrium states $(r^{(k)}, \theta^{(k)}_1, \theta^{(k)}_1, \theta^{(k)}_1 - \psi^{(k)})$, where $-\pi < \theta^{(k)}_1 \leq +\pi$, and $\psi^{(k)}$ is determined by the equilibrium state coordinate $O_k(r^{(k)}, \psi^{(k)})$ of the system \eqref{eq12} describing the chimera in the absence of an external force.

Let us consider for fixed control parameters of the system \eqref{eq1} the region of existence of the equilibrium state $(r^{(k)}, \theta_1^{(k)}, \theta_2^{(k)}, \theta_2^{(k)})$ on the plane of parameters $(\Omega, \varepsilon)$ characterizing the external force. As is known from the synchronization theory \cite{15}, the resulting region will be called the Arnold tongue.
At the base of this region (at $\varepsilon=0$) there will be one point which corresponds to the equilibrium state $(r^{(k)}, \psi^{(k)})$ of the system \eqref{eq12}. For any fixed $\varepsilon \neq 0$, the region will be bounded in frequency by the external force $\Omega$: $\Omega_1(\varepsilon) < \Omega < \Omega_2(\varepsilon)$, these values $\Omega_1(\varepsilon),\Omega_2(\varepsilon)$ will define the boundaries of the Arnold tongue, also called the frequency capture region.

Examples of Arnold tongues are shown in Fig. ~\ref{ris:arn02},~\ref{ris:arn028},~\ref{ris:arn035} (details of the dynamic modes realized within Arnold tongues are described later in Section~\ref{sec:3}). We note that each of the figures indicates two frequency-capturing regions that exist at small values of the $\varepsilon$ parameter. One corresponds to the $O_1$ chimera mode located at the base of the tongue (right) and the other to the $O_2$ chimera mode (left). This is due to the fact that these two frequency capture regions can transition into each other at some non-zero values of the external force amplitude.

In addition, we have to analyze the stability of the found solutions of the system. For this purpose, we linearize the system~\eqref{eq10} and solve the corresponding eigenvalue problem. Thus, the stability of the found structures will be determined by the set of eigenvalues $(\lambda_1, \lambda_2, \lambda_3, \lambda_4)$. The stable solution is defined by the condition $\text{Re} \lambda_i < 0$ $(i = 1,\dots, 4)$. The bifurcation values of the parameters are determined by the equality $\text{Re} \text{Re} \lambda_i = 0$, at least for one value of $i$.

\section{Effects of periodic force on chimera states}
\label{sec:3}
\subsection{Synchronization of a chimera state}
\begin{figure}[!ht]
	\centering{\includegraphics[width=1\linewidth]{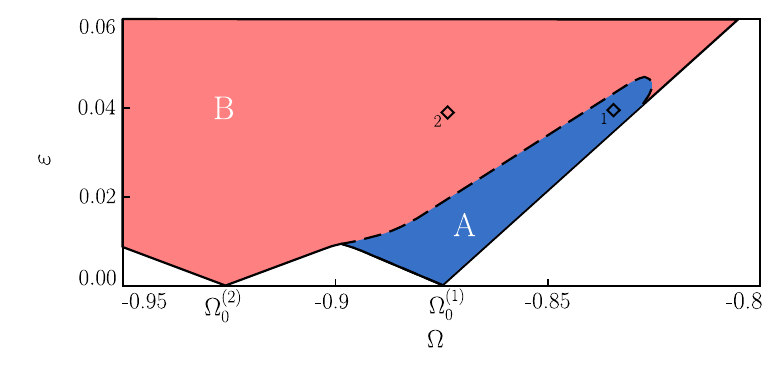}}
	\caption{Frequency locking region (Arnold tongues) of chimera state corresponding to the equilibrium states $O_1$ (with synchronous cluster rotation frequency $\Omega_0^{(1)}$) and $O_2$ (with synchronous cluster rotation frequency $\Omega_0^{(2)}$) of the system~\eqref{eq12} at $\alpha={\pi}/{2}-0. 1$, $\mu_1=0.6$, $\mu_2 = 0.4$. The area $A$, 
		highlighted in blue, denotes stable chimera modes; region $B$, highlighted in red, denotes unstable chimera modes. The black dashed line corresponds to the Andronov--Hopf bifurcation curve for the $O_1$ equilibrium state. Point $1$: $\Omega=-0.8252$, $\varepsilon=0.0409$; point $2$: $\Omega=-0.8431$, $\varepsilon=0.0409$.}
	\label{ris:arn02}
\end{figure}

\begin{figure}[!ht]
	\centering{\includegraphics[width=1\linewidth]{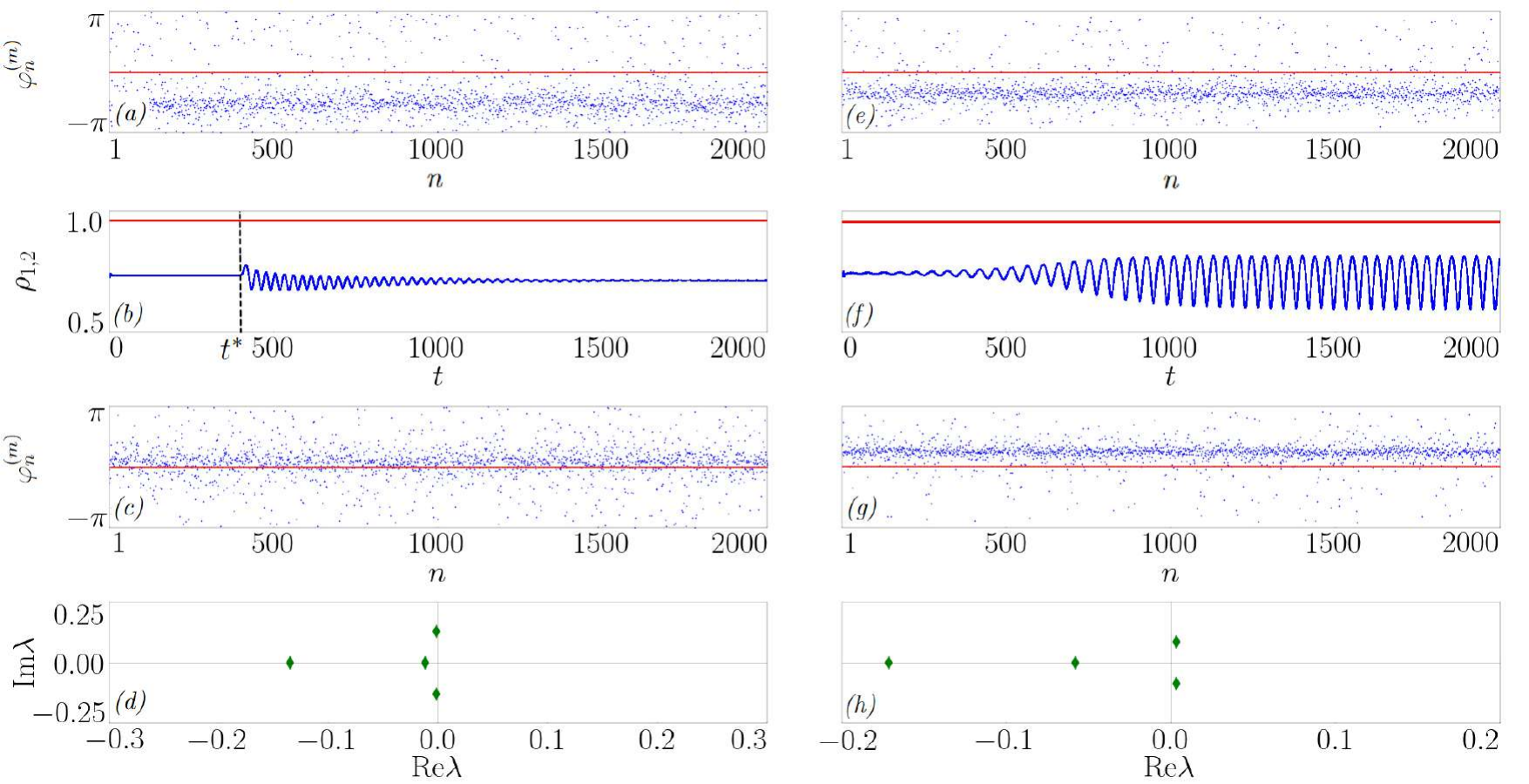}}
	\caption{Results of direct numerical simulation of the phase oscillators~\eqref{eq1} at $N_1=N_2=2000$. (a,c,e,g) Red (blue) markers -- phases of elements of the first (second) ensemble $\varphi_n^{(1)}$ ($\varphi_n^{(2)}$) at initial time moment $t=0$ (fragments (a,e)) and at final time $t=2000$ (fragments (c,g)). (b,f) Dynamics of absolute values of the global order parameters $\rho_1$ (red line) and $\rho_2$ (red line). The black dashed line denotes the moment $t=400$ when the external periodic force is switched on. The fluctuations of $\rho_2$ relative to a constant value at the initial stage of calculations (fragment (b)) are due to the consideration of a finite number of elements in the ensembles $N_1$, $N_2$. (d,h) Eigenvalues $\lambda_i$ ($i = 1, 2, 3, 4$) of the equilibrium state $O_1$ in system~\eqref{eq10}. The parameters are $\alpha={\pi}/{2}-0.1$, $\mu_1=0.6$, $\mu_2 = 0.3$ (Fig.~\ref{ris:arn02}). (a--d) $\Omega = -0.8252, \varepsilon = 0.04$ (point $1$ in Fig.~\ref{ris:arn02}); (e--h) $\Omega = -0.83815, \varepsilon = 0.04$ (point $2$ in Fig.~\ref{ris:arn02}).
	}
	\label{ris:time02}
\end{figure}

Let us consider the effect of an external periodic force on the stable stationary chimera (equilibrium state $O_1$ of the system~\eqref{eq13}). Let us illustrate this case for the parameter values $\omega = 0$, $\alpha = \pi / 2 - 0.1$, $\mu_1 = 0.6$, $\mu_2 = 0.4$, $\Omega = -0.8755$, $\varepsilon = 0$. The corresponding frequency capture region is shown in Fig.~\ref{ris:arn02} (right-hand language).
For small amplitudes of the external force $\varepsilon \leq 0.01$ we obtain the standard form of the Arnold tongue on the plane $(\Omega, \varepsilon)$, within which the chimera with the locked frequency is stable (blue region $A$). The left Arnold tongue containing the saddle chimera (the $O_2$ equilibrium state of the $O_2$ system~\eqref{eq13}) at the base contains only unstable chimera modes at small force amplitudes (red region $B$). The dynamics of the stable mode with the locked frequency is depicted in Fig.~\ref{ris:time02} (left column).
Here, the evolution of the system~\eqref{eq1} without the effect of an external force is shown at the initial stage of the calculation at $t<400$. At $\varepsilon=0$ we observe a stable stationary chimera with a constant value of the ensemble order parameters $r_1 = 1$ and $r_2 < 1$. Then at $t=400$ an external periodic force with amplitude $\varepsilon=0.0409$ and frequency $\Omega =-0.8252$ is switched on. In this case, after a short transient period, the system~\eqref{eq13} evolves to another stationary chimera state induced by the external force. Thus, we can talk about the phenomenon of chimera state synchronization by an external periodic force.

When the parameter $\varepsilon$ increases, the pair of complex-conjugate eigen values of the stable stationary equilibrium state $O_1$ of the system~\eqref{eq13} crosses the imaginary axis (Fig.\ref{ris:time02}h). Consequently, a stable breather chimera emerges in the system~\eqref{eq13} from the stationary chimera regime. Fig.~\ref{ris:time02}f shows the process of development of the stationary chimera state instability leading to a breather chimera. With further increase of the external force amplitude, the unstable stationary chimera mode starts to evolve to a fully synchronous state in which the oscillators in both ensembles are fully phase coherent.

\subsection{Stabilization of an unstable chimera}
\begin{figure}[!ht]
	\centering{\includegraphics[width=1\linewidth]{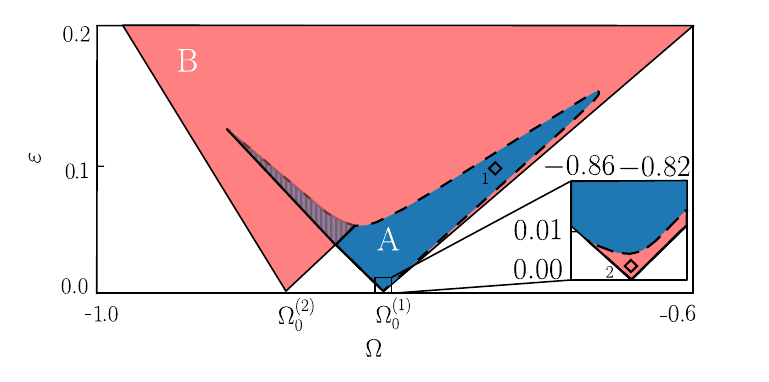}}
	\caption{Same as in Fig.~\ref{ris:arn02}, but for $\alpha= {\pi}/{2}-0.1$, $\mu_1=0.64$, $\mu_2 = 0.36$. The shaded region denotes the overlapping region of the two Arnold tongues. Point $1$: $\Omega=-0.8381$, $\varepsilon=0.0007$; point $2$: $\Omega=-0.7423$, $\varepsilon=0.0997$.}
	\label{ris:arn028}
\end{figure}

\begin{figure}[!ht]
	\centering{\includegraphics[width=1\linewidth]{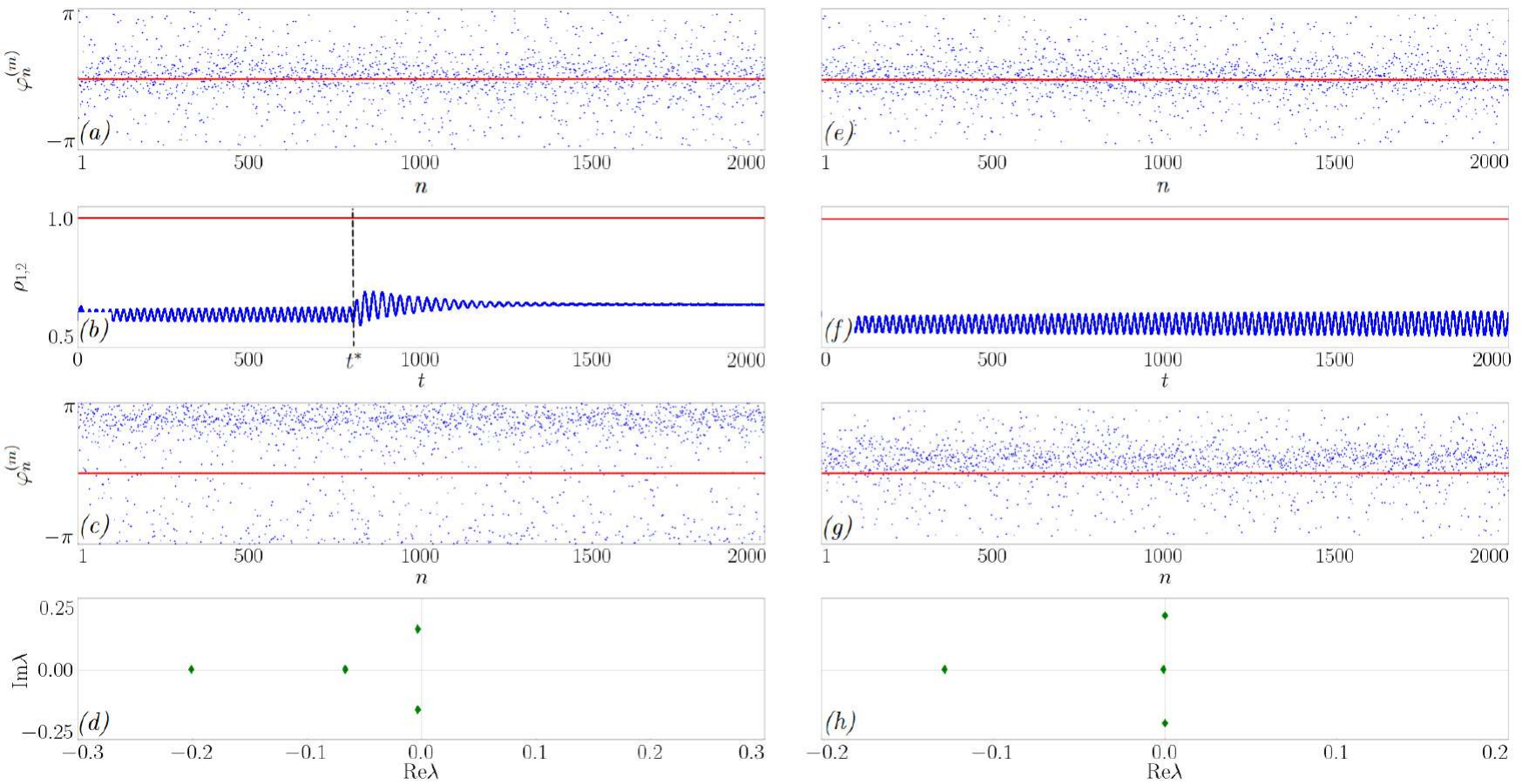}}
	\caption{Same as in Fig. ~\ref{ris:time02}, but for $\alpha= {\pi}/{2}-0.1$, $\mu_1=0.64$, $\mu_2=0.36$ (Fig. ~\ref{ris:arn028}). Parameters: (a--d) $\Omega = -0.7423, \varepsilon = 0.1$ (point $1$ in Fig. ~\ref{ris:arn028}); (e--h) $\Omega = -0.8431, \varepsilon = 0.0007$ (point $2$ in Fig. ~\ref{ris:arn028}).
	}
	\label{ris:time028}
\end{figure}

Let us consider the effect of an external periodic force on the unstable chimera regime (equilibrium state $O_1$ of the system~\eqref{eq13}). Let us focus on several sets of control parameters. At $\omega = 0$, $\alpha = \pi / 2 - 0.1$, $\mu_1 =0.64$, $\mu_2 =0.36$, $\Omega =-0.8381$, $\varepsilon = 0$ in the system~\eqref{eq1} without external force, the stationary chimera is unstable and the oscillator phases evolve to a breather chimera (Fig.~\eqref{ris:phase}b,e). Consequently, unstable modes will be located in the frequency capture region on the $(\Omega, \varepsilon)$ plane (Fig.\ref{ris:arn028}, inset). 
However, when an external force is introduced, the stationary state in the locking region can become stable as a result of the Andronov--Hopf bifurcation. The limit cycle $\Gamma$, which defines the breather chimera, enters the $O_1$ equilibrium state and vanishes. And a stationary chimera regime is observed in the phase oscillator system~\eqref{eq1}. Fig.~\ref{ris:time028}b shows the process of stabilization of the breather chimera due to an external force. 
Initially, at $t<800$, we model the system~\eqref{eq1} without external force with $\varepsilon=0$, and we observe the development of instability of the stationary chimera, resulting in the breather chimera regime. Further, at $t=800$ an external force is introduced, which leads to the disappearance of the breather chimera and the stability of the stationary chimera to which the system~\eqref{eq1} begins to evolve. Further, with increasing coupling strength $\varepsilon$, the stationary chimera becomes unstable again as a result of the Andronov--Hopf bifurcation. When an external influence with characteristics from this region is included, the stable breather from the base of the Arnold tongue evolves to a fully synchronous regime when the phases of the oscillators in both ensembles are fully coherent. 

\begin{figure}[!ht]
	\centering{\includegraphics[width=1\linewidth]{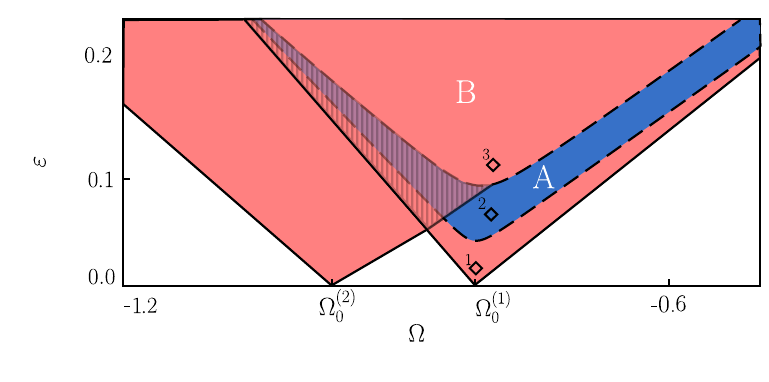}}
	\caption{Same as in Fig.~\ref{ris:arn028}, but for $\alpha= {\pi}/{2}-0.1$, $\mu_1=0.64$, $\mu_2=0.36$. Point $1$: $\Omega=-0.8248$, $\varepsilon=0.0012$; point $2$: $\Omega= -0.8049$, $\varepsilon=0.0808$; point $3$: $\Omega=-0.8001$, $\varepsilon=0.1$.}
	\label{ris:arn035}
\end{figure}

\begin{figure}[!ht]
	\centering{\includegraphics[width=1\linewidth]{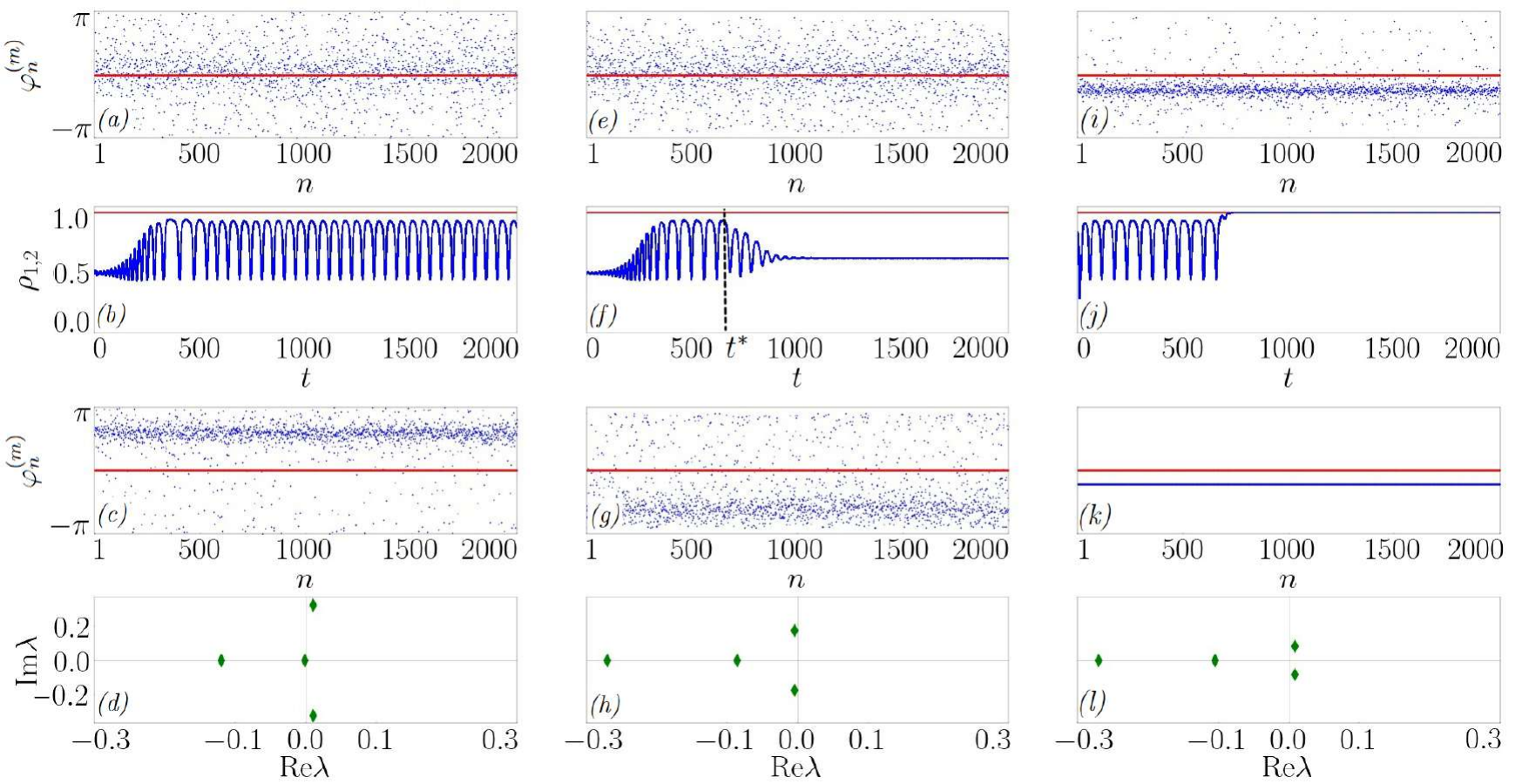}}
	\caption{
		Same as in Fig.~\ref{ris:time02}, but for $\alpha= {\pi}/{2}-0.1$, $\mu_1=0.675$, $\mu_2 = 0.325$ (Fig.~\ref{ris:arn035}). Fragments (i,j,k,l) are similar to fragments (a,b,c,d), respectively. Parameters: (a--d) $\Omega = -0.8248, \varepsilon = 0.0012$ (point $1$ in Fig.~\ref{ris:arn035}); (e--h) $\Omega = -0.8049, \varepsilon = 0. 0808$ (point $2$ in Fig.~\ref{ris:arn035}); (i--l) $\Omega = -0.8001, \varepsilon = 0.1$ (point $3$ in Fig.~\ref{ris:arn035}).
	}
	\label{ris:time035}
\end{figure}

Similarly, given a set of control parameters $\omega = 0$, $\alpha = \pi / 2 - 0.1$, $\mu_1 = 0.675$, $\mu_2 = 0.325$, $\Omega = -0. 8248$, $\varepsilon = 0$ the chimera regime at $\varepsilon=0$ is unstable and evolves to the breather chimera regime with a more nonlinear character of the order parameter oscillations in the partially synchronous ensemble $r(t)$ (Fig.~\ref{eq1}f). Due to the greater degree of the chimera regime instability, the stable equilibrium states region of the reduced system~\eqref{eq13} (region $A$ in Fig.~\ref{ris:arn035}) are further away from the base of the Arnold tongue.
At the same time, the stretching of the area $A$ relative to the boundaries of the frequency capture region is observed, thus stabilization of the chimera regime becomes possible in a wider frequency range of the external force. Note also that due to the increase of the area $A$, the overlap region of the Arnold tongues for the equilibrium states $O_1$ and $O_2$ of the system~\eqref{eq13} (see Fig.~\ref{ris:arn035}) increases significantly. Fig.~\ref{ris:time035} presents the results of numerical experiments on modeling of the system~\eqref{eq1} under several variants of external force.
In the first case (left column of Fig.~\ref{ris:time035}), the development of instability of the stationary chimera is observed, and the system~\ref{eq1} evolves to a breather chimera. The second case (central column of Fig.~\ref{ris:time035}) demonstrates stabilization of the stationary chimera state by an external periodic force applied to the system starting at time $t=750$. In the third case (right column of Fig.~\ref{ris:time035}), the system~\eqref{eq1} evolves to a fully synchronous regime during the development of instability of the stationary chimera.

\section{Conclusions}
\label{sec:4}
In this work, the effect of an external periodic force on chimera solutions in two ensembles of globally coupled identical oscillators was considered. Using the Ott--Antonsen method, dynamical equations were obtained with respect to the global order parameters of the ensembles, characterizing the degree of the elements phases correlation. This allowed us to find chimera modes as equilibrium states of the third order ordinary differential equations. The frequency locking regions (Arnold tongues) of chimera state on the frequency-amplitude parameter plane of the external force were constructed. It is shown that these regions contain subregions with stable and unstable chimera state. In the case of a stable chimera mode realized in the absence of an external force, the possibility of synchronization by a periodic force is demonstrated. In this case, the frequency of the coherent cluster is captured by the frequency of the external force. The effect of a nonstationary chimera (breather) stabilization due to an external periodic force is also shown, when an initially unstable stationary chimera becomes stable. The results obtained in the framework of the reduced model are confirmed by direct numerical modeling of the basic system of phase equations.

\begin{acknowledgments}
The study is supported by the Russian Science Foundation (Sec. II, project 22-12-00348) and the Ministry of Science and Higher Education of Russian Federation  (Sec.III, project FSWR-2023-0034).
\end{acknowledgments}


\begin{thebibliography}{99}%
\bibitem{1} D.M. Abrams, R. Mirollo, S.H. Strogatz,  and D.A. Wiley. {Solvable model for chimera states of coupled oscillators.} \textit{Physical Review Letters}, vol.101(8), 2008.
\bibitem{2} D.M. Abrams and S.H. Strogatz. Chimera states for coupled oscillators. \textit{Phys. Rev. Lett.}, vol.93:174102, Oct 2004.
\bibitem{3} J.A. Acebr\'on, L.L. Bonilla, C.J. P\'erez Vicente, F\'elix Ritort, and Renato Spigler. The Kuramoto model: A simple paradigm for synchronization phenomena. \textit{Reviews of Modern Physics}, vol.77(1):pp.137–185, 2005.
\bibitem{4} T. M. Antonsen, R. T. Faghih, M. Girvan, E. Ott, and J. Platig. External periodic driving of large systems of globally coupled phase oscillators. \textit{Chaos: An Interdisciplinary Journal of Nonlinear Science}, vol.18(3):037112, 09 2008.
\bibitem{5} M.I. Bolotov, L.A. Smirnov, G.V. Osipov, and A. Pikovsky. Locking and regularization of chimeras by periodic forcing. \textit{Phys. Rev. E}, vol.102:042218, Oct 2020
\bibitem{6} P. Clusella, B. Pietras, and E. Montbri\'o. Kuramoto model for populations of quadratic integrate-and-fire neurons with chemical and electrical coupling. \textit{Chaos: An Interdisciplinary Journal of Nonlinear Science}, vol.32(1):013105, 01 2022.
\bibitem{7} J.D. Hart, K. Bansal, T.E. Murphy, and R. Roy. Experimental observation of chimera and cluster states in a minimal globally coupled network. \textit{Chaos}, vol.26(9):094801, 2016.
\bibitem{8} H. Hong. Periodic synchronization and chimera in conformist and contrarian oscillators. \textit{Phys. Rev. E}, vol.89:062924, Jun 2014.
\bibitem{9} Y. Kuramoto. \textit{Chemical Oscillations, Waves, and Turbulence}, vol.19 in \textit{Springer Series in Synergetics}. Springer Berlin Heidelberg, Berlin, Heidelberg, 1984.
\bibitem{10} S. Nkomo, M.R. Tinsley, and K. Showalter. Chimera States in Populations of Nonlocally Coupled Chemical Oscillators. \textit{Physical Review Letters}, vol.110(24):244102, 2013.
\bibitem{11}O. E. Omel’chenko. The mathematics behind chimera states. \textit{Nonlinearity}, vol.31(5):R121–R164, 2018.
\bibitem{12} E. Ott and T.M. Antonsen. Low dimensional behavior of large systems of globally coupled oscillators. \textit{Chaos}, vol.18(3):037113, 2008.
\bibitem{13} M.J. Panaggio and D.M. Abrams. Chimera states: coexistence of coherence and incoherence in networks of coupled oscillators. \textit{Nonlinearity}, vol.28(3):R67–R87, 2015.
\bibitem{14} B. Pietras and A. Daffertshofer. Ott--Antonsen attractiveness for parameter-dependent oscillatory systems. \textit{Chaos}, vol.26(10):103101, 2016.
\bibitem{15} A. Pikovsky, M. Rosenblum, and J. Kurths. \textit{Synchronization : a universal concept in nonlinear sciences}, Cambridge University Press, 2001.
\bibitem{16} A. Pikovsky and M. Rosenblum. Dynamics of globally coupled oscillators: Progress and perspectives. \textit{Chaos}, vol.25(9), 2015.
\bibitem{17} F.A. Rodrigues, T. K. DM. Peron, P. Ji, and J. Kurths. The kuramoto model in complex networks.  \textit{Physics Reports}, vol.610:pp.1–98, 2016.
\bibitem{18} H. Sakaguchi. Cooperative Phenomena in Coupled Oscillator Systems under External Fields. \textit{Progress of Theoretical Physics}, vol.79(1):pp.39–46, 01 1988.
\bibitem{19} M.R. Tinsley, S. Nkomo and K. Showalter. Chimera and phase-cluster states in populations of coupled chemical oscillators. \textit{Nature Physics}, vol.8(9):pp.662–665, 2012.
\bibitem{20} M. Wickramasinghe and I.Z. Kiss. Spatially Organized Dynamical States in Chemical Oscillator Networks: Synchronization, Dynamical Differentiation, and Chimera Patterns. \textit{PLoS ONE}, vol.8(11):e80586, 2013.
\bibitem{21} A.T. Winfree. Biological rhythms and the behavior of populations of coupled oscillators. \textit{Journal of Theoretical Biology}, vol.16(1):pp.15–42, 1967.
\bibitem{22} S. Yamaguchi, H. Isejima, T. Matsuo, R. Okura, K. Yagita, M. Kobayashi, and H. Okamura. Synchronization of cellular clocks in the suprachiasmatic nucleus. \textit{Science}, vol.302(5649):pp.1408–1412, 2003.
\bibitem{23} N. Yao and Z. Zheng. Chimera states in spatiotemporal systems: Theory and Applications. \textit{International Journal of Modern Physics B}, vol.30(07):1630002, 2016.
\end{thebibliography}
\end{document}